\documentstyle[12pt]{article}

\newcommand{\sect}[1]{\setcounter{equation}{0}\section{#1}}

\newcommand{\EQ}{\begin{equation}}
\newcommand{\EN}{\end{equation}}
\newcommand{\bea}{\begin{eqnarray}}
\newcommand{\ena}{\end{eqnarray}}
\newcommand{\vs}[1]{\vspace{#1 mm}}
\newcommand{\hs}[1]{\hspace{#1 mm}}
\renewcommand{\a}{\alpha}
\renewcommand{\b}{\beta}
\renewcommand{\c}{\gamma}

\def\bbox{{\,\lower0.9pt\vbox{\hrule \hbox{\vrule height 0.2 cm
\hskip 0.2 cm \vrule height 0.2 cm}\hrule}\,}}
\newcommand{\dsl}{\pa \kern-0.5em /}

\newcommand{\shalf}{\frac{1}{2}}
\newcommand{\pa}{\partial}
\renewcommand{\t}{\theta}

\newcommand{\nn}{\nonumber\\}
\newcommand{\p}[1]{(\ref{#1})}
\newcommand{\lan}{\langle}
\newcommand{\ran}{\rangle}

\begin{document}

\topmargin 0pt
\oddsidemargin 5mm

\renewcommand{\thefootnote}{\fnsymbol{footnote}}
\begin{titlepage}

\setcounter{page}{0}
\begin{flushright}
OU-HET 288 \\
hep-th/9801023
\end{flushright}

\vs{10}
\begin{center}
{\Large\bf Euclidean Path Integral, D0-Branes and
Schwarzschild Black Holes in Matrix Theory}
\vs{15}

{\large
Nobuyoshi Ohta\footnote{e-mail address: ohta@phys.wani.osaka-u.ac.jp}
and
Jian-Ge Zhou\footnote{jgzhou@phys.wani.osaka-u.ac.jp, JSPS postdoctral fellow}
} \\
\vs{10}
{\em Department of Physics, Osaka University, \\
Toyonaka, Osaka 560, Japan}

\end{center}
\vs{15}
\centerline{{\bf{Abstract}}}
\vs{5}

The partition function in Matrix theory is constructed by Euclidean path
integral method. The D0-branes, which move around in the finite region
with a typical size of Schwarzschild radius, are chosen as the background.
The mass and entropy of the system obtained from the partition function
contain the parameters of the background. After averaging the mass and
entropy over the parameters, we find that they match the properties of 11D
Schwarzschild black holes. The period $\b$ of Euclidean time can be
identified with the reciprocal of the boosted Hawking temperature.
The entropy $S$ is shown to be proportional to the number $N$ of Matrix
theory partons, which is a consequence of the D0-brane background.

\end{titlepage}
\newpage
\renewcommand{\thefootnote}{\arabic{footnote}}
\setcounter{footnote}{0}

\sect{Introduction}

Recently, Banks, Fischler, Klebanov and Susskind found that Matrix
theory~\cite{BFSS} can describe the properties of Schwarzschild black holes,
including the energy-entropy relation, the Hawking temperature and
the physical size up to numerical factors of order
unity~\cite{BFKS1,KS,BFKS2}. Their analysis was done in the limit
that the entropy $S$ of Schwarzschild black holes is proportional to the
number $N$ of Matrix theory partons. In ref.~\cite{HM}, it was pointed out
that the $S\sim N$ limit corresponds to the black hole/black string
transition point, and the thermodynamics of Schwarzschild black holes is
determined by the mean field dynamics of the induced super Yang-Mills zero
modes.

In ref.~\cite{ML}, it was argued that black hole entropy should be
independent of the specifics of the boosting procedure. The
properties of boosted Schwarzschild black holes can be understood in the
framework of an interacting gas of Matrix theory partons. Actually,
in ref.~\cite{BFKS2} the Matrix theory partons -- D0-branes, were treated
as distinguishable (Boltzmannian) particles to fit the relations of black
hole thermodynamics.

To examine the idea of Boltzmann gas of D0-branes, the classical statistical
mechanics of an ensemble of D0-branes in toroidally compactified string
theory was explored in ref.~\cite{LT}, and it was found that the absence
of $1/N!$ factor in the classical Boltzmann partition function is essential
for obtaining the correct black hole thermodynamic functions. In
ref.~\cite{DM}, the concept of infinite statistics was applied to analyse
the properties of Schwarzschild black holes in Matrix theory, and the author
found that D0-branes satisfy quantum infinite statistics.

Other related works on Schwarzschild black holes in Matrix theory
can be found in refs.~\cite{EH,DMRR}. Especially, in \cite{DMRR}
the authors explained that the entropy of boosted $(11-p)$-dimensional
Schwarzschild black holes equal to the entropy of a black
$p$-brane~\cite{BFKS1,KS} is precisely the boosted version of the black
hole/black string transition when the black hole radius $R_S$ becomes
greater than the radius of the compact direction $R$. They argued that
this fact is independent of any property of the 11D M-theory or Matrix
theory, but rather it is a result in any theory which contains General
Relativity in $(d+1)$ dimensions. Furthermore, the evaporation of
Schwarzschild black holes in Matrix theory was discussed in ref.~\cite{BFK},
where the Hawking radiation was realized by emission of small clusters of
D0-branes, and it was found that the rate of the Hawking radiation in Matrix
theory model of Schwarzschild black holes agrees with the semi-classical rate
up to a numerical coefficient of order 1.

On the other hand, Matrix theory purports to be the only candidate for
nonperturbative string theories (M-theory), which probably provides a complete
quantum theory of gravity (even though there is difficulty to find a general
description of Kaluza-Klein compactification of Matrix theory). Thus it is
interesting to calculate the entropy of Schwarzschild black holes from first
principles of Matrix theory. In the context of Einstein gravity, the
Bekenstein-Hawking entropy for Schwarzschild black holes can be derived by
Euclidean path integral approach. The partition function $Z(\b)$ is defined by
\EQ
Z(\b) = {\rm Tr} e^{-\b H},
\label{part1}
\EN
where $H$ is the Hamiltonian operator. The right hand side of eq.~\p{part1}
can be calculated by path integral method, and one has
\EQ
Z(\b) = \int {\cal D}[{\rm path}] e^{-L_E},
\label{part2}
\EN
where $L_E$ denotes the ``Euclidean action", and the integral is taken
over all Euclidean paths which are periodic in Euclidean time with period
$\b$.

Usually one evaluates $Z(\b)$ by expanding around a minimum of $L_E$ and
calculating the contribution to $Z(\b)$ in the one-loop approximation.
In the case of Schwarzschild black hole background, Gibbons and
Hawking~\cite{GH} calculated the path integral in the tree-level approximation,
and found that the entropy derived from the partition function in this
approximation is precisely the Bekenstein-Hawking entropy.
The following questions then naturally arise: Is it possible to calculate the 
Bekenstein-Hawking entropy for Schwarzschild black holes by Euclidean path 
integral method in Matrix theory? What is the proper background for 
Schwarzschild black holes in Matrix theory?

In the present paper, the partition function $Z(\b)$ in Matrix theory is
constructed by Euclidean path integral method. Since the works in
refs.~\cite{BFKS1}-\cite{EH} strongly suggest that Schwarzschild black
holes consist of D0-brane gas, as the first simple exploration, we consider
D0-branes for our background. To compute the partition function $Z(\b)$,
we expand the Lagrangian of Matrix theory to quadratic order in the
fluctuation around the background, and integrate out the off-diagonal
matrix elements which correspond to the degrees of freedom of
the virtual strings stretched among different D0-branes. From the resulting
partition function $Z(\b)$, we read off the mass and entropy of the system,
which contain the parameters of the background. Since the typical size of the
background should be smaller than or equal to Schwarzschild
radius~\cite{BFKS2}, the average values of the parameters of the background
can be obtained. After using these average values of the parameters of the
background in the mass and entropy of the system, we find that they
match the properties of 11D Schwarzschild black holes. From the consistency
of our formalism, we show that the period $\b$ of Euclidean time can be
identified with the reciprocal of the boosted Hawking temperature.

In refs.~\cite{BFKS1,KS,BFKS2}, the analysis was done in the limit
that the entropy $S$ of Schwarzschild black holes is proportional to
the number $N$ of Matrix theory partons. In the present case, the entropy
of the system is derived from the partition function $Z(\b)$ without any
extra assumption, and the $N\sim S$ limit is found to be a consequence of
the fact that D0-branes are exploited as the background. Thus it is
consistent with the picture that D0-brane gas can describe the properties
of 11D Schwarzschild black hole states in the region $S\sim N$.

The layout of the paper is as follows. In the next section we consider the
D0-branes as the background, discuss the restrictions on its parameters, and
construct the partition function $Z(\b)$ in Matrix theory by Euclidean path
integral method. In sect.~3 we calculate the mass, entropy, temperature and
the typical size of the system from the resulting partition function $Z(\b)$
and the consistency of our model. In the derivation, we do not refer to any
information about Schwarzschild black holes except that we require
the typical size of the system to be of order of Schwarzschild radius $R_S$.
As a result, we show that the D0-brane background can be interpreted as
11D Schwarzschild black hole states in Matrix theory. Finally in sect.~4,
we present our discussions of some related issues.

\sect{Construction of partition function $Z(\b)$ in Matrix theory}

In ref.~\cite{BFSS}, the authors proposed that M-theory in the infinite
momentum frame is described as a system of $N\to \infty$ ``partons'',
represented by D0-branes as the carriers of longitudinal momentum.
However, as discussed in ref.~\cite{SBS}, Matrix theory is best thought of
as the Discretized Light-Cone Quantization (DLCQ) of M-theory, {\it i.e.},
compactification on a light-like circle of radius $R$. The dynamics is
dictated by the quantum mechanical Lagrangian with $U(N)$ gauge
symmetry~\cite{BFSS}
\bea
{\cal L} = \frac{1}{2g} {\rm Tr}\left\{ D_0 X^i D_0 X^i + \shalf [X^i,X^j]^2
 + i \t^\dagger D_0 \t + \t^\dagger \c_i [X^i,\t] \right\},
\label{lag0}
\ena
where we have set the string scale $\l_s=1$, and $X^i(i=1, \cdots,9)$ and
$\t$ are bosonic and fermionic hermitian $N\times N$ matrices, respectively.
Since the above Lagrangian possesses gauge symmetry, to calculate the
partition function $Z(\b)$ we have to add gauge fixing and corresponding
ghost terms~\cite{DKPS,LM}. Following ref.~\cite{BFSS}, let us rewrite
the gauge fixed Lagrangian in the unit of 11D Planck length
$l_P$~\cite{DKPS,LM}:
\bea
{\cal L}_T &=& {\rm Tr}\left\{ \frac{1}{2R} D_0 X^i D_0 X^i
 + \frac{R}{4 l_P^6} [X^i,X^j]^2 + i \t^\dagger D_0 \t \right. \nn
&& \left.\hs{5} + \frac{R}{l_P^3}\t^\dagger \c_i [X^i,\t]
 - \frac{1}{2R}({\bar D}^\mu A_\mu)^2 + {\cal L}_g  \right\},
\label{lag1}
\ena
where ${\bar D}^\mu A_\mu = \pa^\mu A_\mu - \frac{iR}{l_P^3}
[{\bar X}^\mu,A_\mu]$, $A^i=X^i$, ${\bar X}^\mu$ is the expectation value
of $A^\mu$, ${\cal L}_g$ is the ghost term and $R$ is the compactification
radius of the light-like coordinate $X^-$~\cite{SBS}.

As usual, we choose ${\bar X}^0=0$ and ${\bar X}^i$ to satisfy the equations
of motion. We then expand \p{lag1} to quadratic order in the fluctuations
around the background fields $X^i={\bar X}^i + \phi^i, A^0 = \phi^0$~\cite{LM}:
\bea
{\cal L}_T &=& {\rm Tr}\left\{ \frac{1}{2R} ({\dot{\bar X^i}})^2
 + \frac{R}{4 l_P^6} [{\bar X}^i,{\bar X}^j]^2
 + \frac{1}{2R} \left[(\pa_0 \phi^i)^2 - (\pa_0 \phi^0)^2 \right]
- \frac{2i}{l_P^3} {\dot{\bar X^i}} [\phi^0,\phi^i] \right. \nn
&&  + \frac{R}{2 l_P^6} \left( [{\bar X}^i,\phi^j]^2
 + [{\bar X}^i,\phi^j] [\phi^i,{\bar X}^j] + [{\bar X}^i,\phi^i]^2
 - [\phi^0,{\bar X}^i]^2 + [{\bar X}^i,{\bar X}^j][\phi^i,\phi^j] \right) \nn
&& \left. + \frac{1}{2R} \pa_0 C^* \pa_0 C
 + \frac{R}{2 l_P^6} [C^*,{\bar X}^i] [{\bar X}^i,C]
 + i \t^\dagger \pa_0 \t + \frac{R}{l_P^3}\t^\dagger \c_i [X^i,\t] \right\},
\label{lag2}
\ena
where the first two terms are the classical parts corresponding to the
tree-level results, and the rests are quadratic in fluctuations contributing
at the one-loop level.

Now we consider the proper background ${\bar X}^i$, which can capture
the essential physics of Schwarzschild black holes. From previous discussions
in~\cite{BFKS1}-\cite{EH}, we know that Schwarzschild black holes consist
of D0-brane gas interacting via the long range static forces~\cite{DKPS}.
To describe the dilute D0-brane gas, the D0-branes should be far apart from
each other. For the first simple exploration, we choose $N$ D0-branes
as the background. For simplicity, we assume that they are moving around
in one direction:\footnote{Physical results are essentially the same if
the D0-branes are moving in various directions.}
\bea
{\bar X}^1 &=& {\rm diag.} ( v_1 t, v_2 t, \cdots, v_k t, \cdots,
 v_l t, \cdots, v_N t), \nn
{\bar X}^2 &=& {\rm diag.} ( b_1, b_2, \cdots, b_k, \cdots, b_l,
 \cdots, b_N), \nn
{\bar X}^i &=& 0, \hs{5} 3 \leq i \leq 9.
\label{back}
\ena

Since we work in Euclidean time with period $\b$, the separations between
$k$-th and $l$-th D0-branes in ${\bar X}^1,{\bar X}^2$ directions are
given by $v_{kl} \b$ and $b_{kl}$ with
\bea
v_{kl} = v_k - v_l; \;\; b_{kl} = b_k - b_l.
\ena
What is the restriction on the parameters of the background \p{back}?
In order to describe the properties of Schwarzschild black holes
in Matrix theory, the background should satisfy a number of
properties~\cite{BFKS2}. For example, the typical size of the background
should be of the magnitude of the order of Schwarzschild radius $R_S$ at
least in the sense of the average, which imposes the following restriction
on the parameters of the background for $|b_{kl}|>> 1$ and $|v_{kl}|<<1$:
\bea
\lan |b_{kl}| \ran \sim
\b \lan |v_{kl}| \ran \sim
\b \lan |v_{k}| \ran \sim R_S,
\label{order1}
\ena
where $\lan \; \ran$ denotes average over the parameters.
Eqs.~(\ref{order1}) indicates that D0-branes move around in the finite
region with a typical size of Schwarzschild radius $R_S$.
Furthermore, the partons saturate the uncertainty bound~\cite{BFKS2,HM}:
\bea
R_S \frac{\lan |v_{k}| \ran}{R} \sim
R_S \frac{\lan |v_{kl}| \ran}{R} \sim 1,
\ena
{\it i.e.}
\bea
\lan |v_{k}| \ran \sim \lan |v_{kl}| \ran \sim \frac{R}{R_S}.
\label{order2}
\ena
Combining \p{order1} and \p{order2}, we find
\bea
\b \sim \frac{R_S^2}{R}.
\label{order3}
\ena
Here we should note that eq.~\p{order3} is obtained from the consistency
of the model, and that we have not used any datum of Schwarzschild black
holes, such as Hawking temperature. As we will see below, the period
$\b$ of Euclidean time can be identified with the reciprocal of
the boosted Hawking temperature, which gives another evidence that
the present model indeed describes the boosted 11D Schwarzschild black holes.

To perform one-loop calculation of the partition function $Z(\b)$, we take
the following form for the fluctuation fields $\phi^\mu,\t$ and $C^\a$:
\bea
\phi^\mu = \left(
\begin{array}{cccccccccc}
0 & a_{12}^\mu & a_{13}^\mu & & \cdots & & & a_{1,N-1}^\mu & a_{1,N}^\mu \\
a_{12}^{\mu\dagger} & 0 & a_{23}^\mu & & \cdots & & & a_{2,N-1}^\mu
 & a_{2,N}^\mu \\
 & & & & \cdots & & & & \\
a_{1m}^{\mu\dagger} & a_{2m}^{\mu\dagger} & \cdots & a_{m-1,m}^{\mu\dagger}
& 0 & a_{m,m+1}^\mu & \cdots & a_{m,N-1}^\mu & a_{m,N}^\mu \\
 & & & & \cdots & & & & \\
a_{1N}^{\mu\dagger} & a_{2N}^{\mu\dagger} & & & \cdots & &
 & a_{N-1,N}^{\mu\dagger} & 0
\end{array}
\right),
\label{flu1}
\ena
where $\mu=t,1,\cdots,9$ and $\phi^t=i\phi^0$, and
\bea
\t = \left(
\begin{array}{cccccccccc}
0 & \psi_{12} & \psi_{13} & & \cdots & & & \psi_{1,N-1} & \psi_{1,N} \\
\psi_{12}^{\dagger} & 0 & \psi_{23} & & \cdots & & & \psi_{2,N-1}
 & \psi_{2,N} \\
 & & & & \cdots & & & & \\
\psi_{1m}^{\dagger} & \psi_{2m}^{\dagger} & \cdots & \psi_{m-1,m}^{\dagger}
& 0 & \psi_{m,m+1} & \cdots & \psi_{m,N-1} & \psi_{m,N} \\
 & & & & \cdots & & & & \\
\psi_{1N}^{\dagger} & \psi_{2N}^{\dagger} & & & \cdots & &
 & \psi_{N-1,N}^\dagger & 0
\end{array}
\right), \nn
C^\a = \left(
\begin{array}{cccccccccc}
0 & C_{12}^\a & C_{13}^\a & & \cdots & & & C_{1,N-1}^\a & C_{1,N}^\a \\
C_{12}^{\a\dagger} & 0 & C_{23}^\a & & \cdots & & & C_{2,N-1}^\a
 & C_{2,N}^\a \\
 & & & & \cdots & & & & \\
C_{1m}^{\a\dagger} & C_{2m}^{\a\dagger} & \cdots & C_{m-1,m}^{\a\dagger}
& 0 & C_{m,m+1}^\a & \cdots & C_{m,N-1}^\a & C_{m,N}^\a \\
 & & & & \cdots & & & & \\
C_{1N}^{\a\dagger} & C_{2N}^{\a\dagger} & & & \cdots & &
 & C_{N-1,N}^{\a\dagger} & 0
\end{array}
\right),
\label{flu2}
\ena
where $\a=1,2$. The diagonal elements of the fluctuation fields are set to
zero because their contribution to the partition function can be ignored in
the one-loop approximation on the background~\p{back}.

Inserting \p{back}, \p{flu1} and \p{flu2} into \p{lag2}, and changing the
time $t\to it$ and velocity $v\to -iv$, we have
\bea
{\cal L}_E &=& \sum_{l=1}^N \frac{v_l^2}{2R} + \frac{1}{R}\left\{ \sum_{k<l}^N
(a_{kl}^{t\dagger}, a_{kl}^{1\dagger}) \left(
\begin{array}{cc}
-\pa_t^2 + \frac{R^2(b_{kl}^2 + v_{kl}^2 t^2)}{l_P^6}
 & -\frac{2iRv_{kl}}{l_P^3} \\
\frac{2iRv_{kl}}{l_P^3}
 & -\pa_t^2 + \frac{R^2(b_{kl}^2 + v_{kl}^2 t^2)}{l_P^6}
\end{array} \right) \left(
\begin{array}{c}
a_{kl}^t \\ a_{kl}^1 \end{array}
\right) \right. \nn
&& + \sum_{i=2}^9 a_{kl}^{i\dagger} \left[
 -\pa_t^2 + \frac{R^2(b_{kl}^2 + v_{kl}^2 t^2)}{l_P^6} \right] a_{kl}^i
+ \sum_{\a=1}^2 C_{kl}^{\a\dagger} \left[
 -\pa_t^2 + \frac{R^2(b_{kl}^2 + v_{kl}^2 t^2)}{l_P^6} \right] C_{kl}^\a \nn
&& \left. + \psi_{kl}^\dagger \left[
 \pa_t - \frac{R (b_{kl} \c_2 + v_{kl} t\c_1)}{l_P^3} \right] \psi_{kl}
\right\},
\label{lagq}
\ena
where $t$ is the Euclidean time. Note that eq.~\p{lagq} shows that in
the one-loop approximation the elements $a_{kl}^i, C_{kl}^\a$ and
$\psi_{kl}$ decouple from other components $a_{k'l'}^i, C_{k'l'}^\a$
and $\psi_{k'l'}$ ($k \neq k'$ and $l\neq l'$).

According to eq.~\p{part2}, we define the partition function $Z(\b)$ as
\bea
Z(\b) = \int \prod_{k<l}^N \prod_{i=1}^9 \prod_{\a=1}^2
 {\cal D}a_{kl}^i {\cal D}a_{kl}^{i\dagger} {\cal D}a_{kl}^t
 {\cal D}a_{kl}^{t\dagger} {\cal D}C_{kl}^\a {\cal D}C_{kl}^{\a\dagger}
 {\cal D}\psi_{kl} {\cal D}\psi_{kl}^\dagger
 \exp\left[- \int_{-\frac{\b}{2}}^{\frac{\b}{2}} {\cal L}_E dt \right],
\ena
with the (anti-)periodic boundary conditions\footnote{It is customary to
take the range of Euclidean time $t$ from 0 to $\b$. We have chosen it
from $-\frac{\b}{2}$ to $\frac{\b}{2}$ for our convenience. Not only can
this be justified but also does not affect our later estimate of the
order of magnitudes of various physical quantities.}
\bea
\Phi_{kl}\left(-\frac{\b}{2} \right) = \pm \Phi_{kl}\left(\frac{\b}{2} \right),
\Phi_{kl}^\dagger \left(-\frac{\b}{2} \right)
 = \pm \Phi_{kl}^\dagger \left(\frac{\b}{2} \right),
 \ena
where $\Phi$ denotes $a^i,a^t,C^\a$ and $\psi$ and the sign is $+(-)$ for
bosons and ghosts (fermions).

After integrating over $\Phi_{kl}$ and $\Phi_{kl}^\dagger$,
we have~\cite{DKPS}-\cite{TP}
\EQ
Z(\b) = e^{-(\Gamma_0 + \Gamma_1)},
\label{part}
\EN
where $\Gamma_0$ is the contribution from the tree-level and $\Gamma_1$
is that from the one-loop. They can be expressed as
\EQ
\Gamma_0 = \b \sum_{l=1}^N \frac{v_l^2}{2R},
\label{tree}
\EN
and
\bea
\Gamma_1 &=& - \sum_{k<l}^N \ln \left\{
{\rm det}^{-6}\left[ -\pa_t^2 + \frac{R^2(b_{kl}^2 + v_{kl}^2 t^2)}{l_P^6}
 \right]
{\rm det}^{-1}\left[ -\pa_t^2 + \frac{R^2(b_{kl}^2 + v_{kl}^2 t^2)}{l_P^6}
 + \frac{2R v_{kl}}{l_P^3} \right]
\right. \nn
&& \times
{\rm det}^{-1}\left[ -\pa_t^2 + \frac{R^2(b_{kl}^2 + v_{kl}^2 t^2)}{l_P^6}
 - \frac{2R v_{kl}}{l_P^3} \right]
{\rm det}^4\left[ -\pa_t^2 + \frac{R^2(b_{kl}^2 + v_{kl}^2 t^2)}{l_P^6}
 + \frac{R v_{kl}}{l_P^3} \right] \nn
&& \left. \times
{\rm det}^4\left[ -\pa_t^2 + \frac{R^2(b_{kl}^2 + v_{kl}^2 t^2)}{l_P^6}
 - \frac{R v_{kl}}{l_P^3} \right] \right\}.
\label{1loop}
\ena
The interval of $t$ is $-\frac{\b}{2} \leq t \leq \frac{\b}{2}$.

The calculation of the above determinants is closely related to that
in ref.~\cite{TP} where the finite time amplitudes in Matrix theory
was discussed. In order to derive the leading order result, it is sufficient
to make the adiabatic approximation; namely, we retain only the contribution
from the ground state. Thus we make the approximation
\EQ
{\rm det}\left[-\pa_t^2+\omega^2(t)\right]
 \simeq \exp\left[ \int_{-\frac{\b}{2}}^{\frac{\b}{2}} \omega(t) dt \right].
\label{app}
\EN
We consider large Schwarzschild black holes with $b_{kl} >> 1$ and
$v_{kl}<<1$. The leading term for $\Gamma_1$ can then be written as~\cite{TP}
\EQ
\Gamma_1 = - \frac{15 l_P^9}{16 R^3} \sum_{k<l}^N
\int_{-\frac{\b}{2}}^{\frac{\b}{2}} dt \frac{v_{kl}^4}{(b_{kl}^2
 + v_{kl}^2 t^2)^{7/2}}.
\label{1loop1}
\EN
Substituting \p{tree} and \p{1loop1} into \p{part}, the partition
function becomes
\EQ
Z(\b) = \exp\left\{ -\left[ \b \sum_{l=1}^N \frac{v_l^2}{2R}
- \frac{15 G_{11}}{16 R^3} \sum_{k<l}^N
 \int_{-\frac{\b}{2}}^{\frac{\b}{2}} dt \frac{v_{kl}^4}
 {(b_{kl}^2 + v_{kl}^2 t^2)^{7/2}} \right] \right\},
\label{partf}
\EN
where $G_{11}=l_P^9$.

Up to now, we have arrived at the first goal, that is, we have successfully
constructed the partition function $Z(\b)$ in the one-loop approximation in
Matrix theory. In the next section, we will use it to calculate
the energy and entropy of the system, and compare them with the properties
of 11D Schwarzschild black holes.

\sect{The energy, entropy and temperature of the system}

Let us first consider the energy of the system. From eq.~\p{partf}, we can
express it as
\bea
{\tilde E}_e &=& -\frac{\pa \ln Z(\b)}{\pa \b} \nn
&=& \sum_{l=1}^N \frac{v_l^2}{2R} - \frac{15 G_{11}}{16 R^3} \sum_{k<l}^N
 \frac{v_{kl}^4}{(b_{kl}^2 + v_{kl}^2 \b^2/4)^{7/2}},
\label{free}
\ena
where the tilde means that the energy has not been averaged over
the parameters. Here we point out that in deriving \p{free}, we have not
made any approximation in \p{partf}. So we use the subscript ``$e$'' to
denote the exact calculation.\footnote{By ``exact", we mean that no
approximation is made to eq.~\p{partf}, which was obtained in the
approximation \p{app} and \p{1loop1}. This is to be contrasted with the
perturbative result~\p{free2} below.}

Now let us average the energy ${\tilde E}_e$ over the parameters by
using eqs.~\p{order1}-\p{order3}:
\bea
E_e &=& \lan {\tilde E}_e \ran \nn
&=& \sum_{l=1}^N \frac{\lan v_l^2\ran}{2R} - \frac{15 G_{11}}{16 R^3}
 \sum_{k<l}^N \frac{\lan v_{kl}^4\ran}{\lan (b_{kl}^2
 + v_{kl}^2 \b^2/4 )^{7/2}\ran} \nn
&\sim& \frac{NR}{2R_S^2} - \shalf \left(\frac{4}{5}\right)^{\frac{7}{2}}
 \frac{15 G_{11}RN^2}{16 R_S^{11}}.
\label{ee}
\ena

On the other hand, since $v_{kl}^2/b_{kl}^2 << 1$, we expand \p{partf}
in the power of $v_{kl}^2/b_{kl}^2$ to get
\bea
-\ln Z(\b) \simeq \b \sum_{l=1}^N \frac{v_l^2}{2R}
 - \frac{15 G_{11}\b}{16 R^3} \sum_{k<l}^N \frac{v_{kl}^4}{b_{kl}^7}
 + \frac{15 G_{11}\b^3}{16 R^3} \sum_{k<l}^N \frac{7v_{kl}^6}{24b_{kl}^9}.
\label{free1}
\ena
From eq.~\p{free1}, the energy can be read off as
\bea
{\tilde E}_p &=& -\frac{\pa\ln Z{(\b)}}{\pa\b} \nn
&=& \sum_{l=1}^N \frac{v_l^2}{2R}
 - \frac{15 G_{11}}{16 R^3} \sum_{k<l}^N \frac{v_{kl}^4}{b_{kl}^7}
 + \frac{15 G_{11}\b^2}{16 R^3} \sum_{k<l}^N \frac{7v_{kl}^6}{8b_{kl}^9},
\label{free2}
\ena
where the subscript ``$p$'' denotes that the energy is obtained from the
perturbative $Z(\b)$, in contrast to that in eq.~\p{free}. We then have
\bea
E_p &=& \lan {\tilde E}_p\ran \nn
&\sim& \frac{NR}{2R_S^2} - \frac{1}{16} \frac{15 G_{11}RN^2}{16 R_S^{11}}.
\label{free3}
\ena
Comparing \p{free3} with \p{ee}, we find that $\b^2 \lan v_{kl}^2/b_{kl}^2
\ran \sim 1$ and higher order terms have the same order
of magnitudes as the lower terms, but they just affect the coefficient of
the second term in \p{free3}. Thus in order to estimate the order of
magnitudes and also dependence on various physical quantities (up to
numerical coefficients of order unity), we can use eq.~\p{free1} to
calculate the energy and entropy of the system. From eq.~\p{free1},
the Helmholtz free energy is given by
\bea
{\tilde F} &=& -\frac{1}{\b} \ln Z(\b) \nn
&=& \sum_{l=1}^N \frac{v_l^2}{2R}
 - \frac{15 G_{11}}{16 R^3} \sum_{k<l}^N \frac{v_{kl}^4}{b_{kl}^7}
 + \frac{15 G_{11}\b^2}{16 R^3} \sum_{k<l}^N \frac{7v_{kl}^6}{24b_{kl}^9},
\ena
Then the entropy of the system is
\bea
{\tilde S} &=& \b^2 \frac{\pa {\tilde F}}{\pa \b} \nn
&=& \frac{15 G_{11}\b^3}{8 R^3} \sum_{k<l}^N \frac{7v_{kl}^6}{24b_{kl}^9}.
\ena
After averaging ${\tilde S}$ over the parameters, one has
\EQ
S=\lan{\tilde S}\ran \sim \frac{G_{11}N^2}{R_S^9}.
\label{ent}
\EN
Here we emphasize that the entropy of the system is derived directly from
the partition function $Z(\b)$ defined in sect.~2.

To compare our results with previous estimates in refs.~\cite{BFKS1}-\cite{EH},
we apply the virial theorem to \p{ee} to find
\EQ
R_S \sim (G_{11}N)^{1/9}.
\label{rs}
\EN
Inserting \p{rs} into eqs.~\p{order3}, \p{free3} and \p{ent} yields
\bea
E &\sim& \frac{(G_{11}^{-1/9} N^{8/9})^2 R}{N},
\label{rese} \\
S &\sim& N,
\label{ress} \\
\b &\sim& \frac{(G_{11} N)^{2/9}}{R}.
\label{resb}
\ena
Eq.~\p{ress} is the condition $N\sim S$ assumed
in refs.~\cite{BFKS1,KS,BFKS2}, but here we have derived it
without any extra assumption. It is just a consequence from the fact that
D0-branes are chosen as the background. Since the energy $E$ in Matrix
theory is the light-cone energy related to the mass $M$ of a boosted
object by~\cite{BFKS1}-\cite{EH}
\EQ
E=\frac{M^2 R}{2N},
\EN
the mass of the object is
\EQ
M \sim G_{11}^{-1/9} N^{8/9},
\label{mass}
\EN
and the $\b$ can be interpreted as the reciprocal of the boosted
temperature of the system. This implies that the temperature $T$ in the
rest frame can be related with $\b$ by
\EQ
\frac{1}{\b} = \frac{MR}{N}T \sim \frac{R}{(G_{11} N)^{2/9}}.
\label{temp}
\EN
Substituting \p{mass} into \p{temp}, we have
\EQ
T \sim \frac{1}{(G_{11} N)^{1/9}} \sim \frac{1}{R_S}.
\EN
Since the entropy, which have an interpretation in terms of total number of
states, should not change under the boost, the mass, entropy, temperature
and typical size of the system can be collectively written as
\bea
M &\sim& G_{11}^{-1/9} N^{8/9}, \nn
S &\sim& N, \nn
T &\sim& \frac{1}{R_S}, \nn
R_S &\sim& (G_{11} N)^{1/9}.
\label{res}
\ena

In deriving the result \p{res}, we have not referred to any information
about Schwarzschild black holes except that we assume that the typical size
of the system is of the order of Schwarzschild radius $R_S$. Also our model
does not contain General Relativity apparently; it is just Matrix theory
itself. What we have shown is that we can derive \p{res} solely from the
consistency of the model, including Euclidean path integral, backgrounds
and restriction on its parameters. From eq.~\p{res}, it is easy to see that
\bea
R_S &\sim& (G_{11} M)^{1/8}, \nn
S &\sim& G_{11}^{1/8}M^{9/8} \sim N, \nn
T &\sim& \frac{1}{R_S}.
\label{res1}
\ena
We find that eqs.~\p{res1} are nothing but the thermodynamic functions
of 11D Schwarzschild black holes, which indicates that our model indeed 
describes 11D Schwarzschild black holes. Since the model does not involve
General Relativity manifestly, we conclude that Matrix theory itself
contains 11D Schwarzschild black hole states or their disguise.
The background we choose is D0-branes, which is closely related to D0-brane
gas picture in refs.~\cite{BFKS1}-\cite{EH}; actually it describes
the black hole states in the region $S\sim N$.

\sect{Discussions}

So far we have developed Euclidean path integral formalism in Matrix theory
to construct the partition function $Z(\b)$, from which the energy and
entropy of the system can be read off. In order to calculate the
partition funcution $Z(\b)$ in the one-loop approximation, the background
has been assumed to be D0-branes moving around in the finite region with
a typical size of Schwarzschild radius $R_S$, which can be visualized
as a bound state of a large number of D0-branes. In fact, such a bound
state can be interpreted as an 11D Schwarzschild black hole state in the
language of Matrix theory. As we have shown, with only one assumption
that the typical size of the system is of the order of Schwarzschild radius,
we can derive the mass, entropy and temperature of the system up to
numerical factors of order unity by the consistency of the formalism.
Also we have found that they match the properties of 11D Schwarzschild
black holes. The assumption $S\sim N$ made in \cite{BFKS1,KS,BFKS2}
has been clarified in our model, which is a direct consequence of
D0-brane background. In other words, the D0-brane gas only describes
the properties of 11D Schwarzschild black hole states in the region
$S\sim N$.

It is interesting to compare the present Euclidean path integral approach
with what Gibbons and Hawking proposed in ref.~\cite{GH}, where they used
the metric of Schwarzschild black holes as the background.
The Bekenstein-Hawking entropy was read off in the tree-level approximation
from the partition function in their calculation. However, in Matrix
theory, we have exploited D0-branes as the background, and found that
one-loop corrections are enough to find the correct entropy
(the contribution from tree-level vanishes). It is quite reminiscent of
Enclidean path integral formalism in the calculation of the entropy
of black holes from matter fields~\cite{DO,DK}, where the entropy
only gets contribution from one-loop corrections, but the background was still
chosen as the metric of black holes. Probably, treating D0-branes as the
background is the main new feature of the present Euclidean path integral
method in Matrix theory.

One of the interesting extensions of our formalism is to check whether
it works in various dimensions to describe Schwarzshild black hole states
in the corresponding dimension, which is related to finding a general
descripition of Kaluza-Klein compactification of Matrix
theory~\cite{WT,GRT,CDS}.

In the above, we have learned that 11D Schwarzschild black hole states
in the region $S\sim N$ can be correctly described in Matrix theory by
choosing D0-branes as the background. One may ask what happens in 11D
Schwarzschild black hole states in Matrix theory in the limit $N>>S$.
In ref.~\cite{HM}, it was speculated on that if the transverse size
remains constant under boosts, the partons become denser as $N$ increases
and strongly interacting clusters will form. The interaction within
a cluster should be more ``membrane-like'' than ``graviton-like",
since the commutator term in the Matrix-theory Hamitonian is the membrane
area element. Now it seems that in our formalism we can examine if
the above idea is realized, that is, we can describe 11D Schwarzschild
black hole states in Matrix theory in the limit $N>>S$ with certain
background. If we choose a number of little nuggets of membranes as
the background, one might expect the resulting entropy $S$ will be
much less than $N$, but the mass, entropy, temperature and the typical size
of the system still match the properties of the 11D Schwarzschild black holes.
Work along this line is in progress and we hope to discuss these issues
elsewhere.

\section*{Acknowledgement}

J.-G.Z. would like to thank T. Nakatsu and K. Oda for discussions.
This work was supported in part by Grand-in-aid from the Ministry of
Education, Science, Sports and Culture No. 96208.

\newpage
\newcommand{\NP}[1]{Nucl.\ Phys.\ {\bf #1}}
\newcommand{\AP}[1]{Ann.\ Phys.\ {\bf #1}}
\newcommand{\PL}[1]{Phys.\ Lett.\ {\bf #1}}
\newcommand{\CQG}[1]{Class. Quant. Gravity {\bf #1}}
\newcommand{\NC}[1]{Nuovo Cimento {\bf #1}}
\newcommand{\CMP}[1]{Comm.\ Math.\ Phys.\ {\bf #1}}
\newcommand{\PR}[1]{Phys.\ Rev.\ {\bf #1}}
\newcommand{\PRL}[1]{Phys.\ Rev.\ Lett.\ {\bf #1}}
\newcommand{\PRE}[1]{Phys.\ Rep.\ {\bf #1}}
\newcommand{\PTP}[1]{Prog.\ Theor.\ Phys.\ {\bf #1}}
\newcommand{\PTPS}[1]{Prog.\ Theor.\ Phys.\ Suppl.\ {\bf #1}}
\newcommand{\MPL}[1]{Mod.\ Phys.\ Lett.\ {\bf #1}}
\newcommand{\IJMP}[1]{Int.\ Jour.\ Mod.\ Phys.\ {\bf #1}}
\newcommand{\JP}[1]{Jour.\ Phys.\ {\bf #1}}

\end{document}